\documentclass[aps,prb,twocolumn,epsfig,floats,superscriptaddress]{revtex4}
\usepackage{graphicx,epsf,natbib,amsmath,latexsym,amssymb,multirow}
\usepackage[parfill]{parskip}    		
\usepackage{graphicx}				
\usepackage{float}								
\usepackage{amssymb}
\usepackage{amsmath}
\usepackage[per-mode=fraction]{siunitx}    
\DeclareSIQualifier\bohrB{B}
\DeclareSIUnit\bohrM{\micro \bohrB}

\begin{document}

\title{Tunable line node semimetals}
\author{Michael Phillips and Vivek Aji} 
\affiliation{Department of Physics \& Astronomy, University of California, Riverside CA 92521} 
\begin{abstract}

Weyl semimetals are examples of a new class of topological states of matter which are gapless in the bulk with protected surface states. 
Their low energy sector is characterized by massless chiral fermions which are robust against translationally invariant perturbations. A
variant of these systems have two non-degenerate bands touching along lines rather than points. A proposal to realize such a phase involves
alternating layers of topological insulators and magnetic insulators, where the magnetization lies perpendicular to the symmetry axis of the
heterostructure. The shape, size and the dispersion in the vicinity of the nodal lines varies with the strength of the magnetization, offering
a new knob to control the properties of the system. In this paper we map out the evolution of the nodal lines and the dependence of the conductivity 
on magnetization and identify signatures of the low energy sector in quantum oscillation measurements.

\end{abstract}

\maketitle
\renewcommand{\thefootnote}{\roman{footnote}}

\section{Introduction}

The discovery of topological states of matter has brought a revolution in solid state physics. They provide impetus to develop new methodologies to find
and characterize them, both theoretically and experimentally, and have opened new directions for technological innovation. Topological insulators in two and three
dimensions which are gapped in the bulk and have surface states have already been realized. An important insight gained from these developments is the prominent role
played by spin-orbit interaction in stabilizing such nontrivial phases\cite{haldane, kane,zhangh,konig,kane,teo, zhangrevmodphys,fu, xu, fu1, roy, jemoore}. A consequence of this line of investigation lead to the remarkable conjecture that some pyrochlore iridates, 
which possess all the necessary ingredients, such as large atomic number, localized moments and moderate correlation, are in a semimetallic phase where two nondegenrate bands
touch at an even number of points in the Brillouin zone\cite{xwan}. Moreover these nodes in the energy landscape are at the chemical potential and the low energy sector is described in terms of chiral massless electrons in three dimensions, which were first discovered by Hermann Weyl\cite{weyl}.

While pyrochlore iridates have yet to be unambiguously shown to house such a phase, a number of proposals have appeared in the literature that have the potential to do so. Balents and Burkov\cite{bb} showed that a heterostructure
made up of alternating layers of magnetically doped topological insulator and normal insulator had
Weyl fermions in its low energy sector. An alternate route is to find materials which have a Dirac
dispersion in three dimensions and lift the spin degeneracy by breaking either time reversal or inversion. 
Angle resolved photoemission spectroscopy measurements on Na$_{3}$Bi and Cd$_{3}$As$_{2}$ have
provided the first evidence for the existence of massless Dirac fermions\cite{zwangtop,zkl, www,sb,neupane}. The latter also breaks inversion 
and has the potential to be a Weyl semimetal, but the data lacks the resolution to verify the claim.

Here we focus on a variant of the heterostructure where a line is obtained instead of point nodes\cite{bhb}. This requires the magnetization
of either the magnetically doped topological insulator or that of the ferromagnetic insulator to be perpendicular rather than
parallel to the symmetry axis of the device. The shape anisotropy of the device naturally favors such a geometry. Alternatively one can use an antiferromagnetic
insulator with a suitable choice of terminating surfaces to provide the uniform exchange field needed. This construction has the advantage of the ability to
tune the magnetization by varying temperature. This provides a knob to manipulate the response of the device and access the interesting semimetallic phase. The main motivation of the study is that for line nodes the size, shape and density of states all depend on the magnetization. This is in contrast with the nodal semimetal where only the distance between the nodes depends on mangetization. The evolution of the low energy sector, as well as its consequence on thermodynamic and transport properties, as a function of magnetization is explored in this paper. 

\section{Model} \label{sec:initial}

As described by A. A. Burkov, M. D. Hook and Leon Balents\cite{bhb}, a simple way to construct a Weyl semimetal is to arrange alternating layers of topological insulator (TI) and normal insulator (NI). This setup leads to a Weyl semimetal dispersion containing the minimum of two nodes, provided time reversal symmetry is broken. To achieve this, the addition of magnetic impurities in each TI layer was proposed with magnetization along the $z$-direction -- orthogonal to each layer, along the direction of growth.  The two materials are set up such that each pair of layers (TI + NI) add up to a thickness $d$.  

The full 2D Hamiltonian in terms of the momentum ${ \vec { k }  }_{\bot}=k_x \hat{x} + k_y \hat{y}$ describing this multilayered structure (using the notation and formalism in [\onlinecite{bhb}]) is 
\begin{align}  
{H} {=  }&{ \sum _{ { \vec { k }  }_{ \bot  },ij }{ { c }_{ { \vec { k }  }_{ \bot  }i }^{ \dagger }{ c }_{ { \vec { k }  }_{ \bot  }j } [ { v }_{ F }{ \tau  }^{ z }(\hat { z } \times \vec { \sigma  } )\cdot { \vec { k }  }_{ \bot  }{ \delta  }_{ ij }  +m{ \sigma  }^{ z }{ \delta  }_{ ij }  }   }     \notag  \\
&{+ { \Delta  }_{ S }{ \tau  }^{ x }{ \delta  }_{ ij } + \frac { 1 }{ 2 } { \Delta  }_{ D }({ \tau  }^{ + }{ \delta  }_{ j,i+1 } + { \tau }^{ - }{ \delta  }_{ j,i-1 }) ]  .}   \label{eq:hamil1}
\end{align}

The first term describes the top and bottom states of a single TI layer (with $\hbar=1$).  The second term describes the spin splitting, resulting from magnetization in the $z$-direction.  The remaining terms describe tunneling within an individual TI layer (the $\Delta_{S}$ term), and between neighboring TI layers (the $\Delta_{D}$ terms).   Without loss of generality one can set $\Delta_{S}, \Delta_{D}>0$.

The eigenvalues for this Hamiltonian lead to the dispersion
\begin{equation} \label{eq:dis1}
{ \varepsilon  }_{ \pm  }^{ 2 }={ v }_{ F }^{ 2 }\left| { \kappa  }_{ \bot  } \right| ^{ 2 }+(m\pm \left| \Delta ({ k }_{ z }) \right| )^{ 2 },
\end{equation}
 where $ { \Delta  }({ k }_{ z }) = {\Delta}_{S}+{\Delta}_{D}  { e }^{  i   { k }_{ z }d } $ and $ { \kappa  }_{ \bot  } =  { k }_{ y }+i{ k }_{ x } $. There are a pair of non-degenerate nodes located at $ k_x=k_y=0, k_z= \frac{\pi}{d} \pm k_0 $ where 
\begin{align} \label{eq:k0bb}
k_0=\frac{1}{d}\arccos { \left[ 1-\left( \frac { { m }^{ 2 }-({ \Delta  }_{ S }-{ \Delta  }_{ D })^{ 2 } }{ 2{ \Delta  }_{ S }{ \Delta  }_{ D } }  \right)  \right]  }. &
\end{align}
The nodes exist provided
\begin{equation} \label{eq:cond1} 
 ({ \Delta  }_{ S }-{ \Delta  }_{ D })^{ 2 }<{ m }^{ 2 }<({ \Delta  }_{ S }+{ \Delta  }_{ D })^{ 2 }.  
\end{equation}
Such  Weyl semimetals are expected to display a number of anomalous properties and house novel correlated phases. A variant of this setup is one where the 
axial symmetry is broken in addition to time reversal.  The low energy sector is this case has line nodes and a system that is less studied.

Choosing the magnetization to be along the $x$-axis  modifies the second term in eqn. (\ref{eq:hamil1}).  In practice this can be achieved by replacing the normal insulator with either ferromagnetic insulator, or antiferromagnetic insulator with appropriately chosen terminating surface. The Hamiltonian becomes
$$
{H =   \sum _{ { \vec { k }  }_{ \bot  },ij }{ { c }_{ { \vec { k }  }_{ \bot  }i }^{ \dagger }{ c }_{ { \vec { k }  }_{ \bot  }j } [ { v }_{ F }{ \tau  }^{ z }(\hat { z } \times \vec { \sigma  } )\cdot { \vec { k }  }_{ \bot  }{ \delta  }_{ ij }  +m{ \sigma  }^{ x }{ \delta  }_{ ij }  }   }
$$
\begin{equation} \label{eq:hamil2}
{ +{ \Delta  }_{ S }{ \tau  }^{ x }{ \delta  }_{ ij }+\frac { 1 }{ 2 } { \Delta  }_{ D }({ \tau  }^{ + }{ \delta  }_{ j,i+1 }+{ \tau }^{ - }{ \delta  }_{ j,i-1 }) ]  }.
\end{equation}

The resulting dispersion is
\begin{align} \label{eq:dis2}
{ \varepsilon  }_{ \pm  }^{ 2 }={ v }_{ F }^{ 2 } {k}_{x}^{ 2 }+ \left( m\pm \sqrt{{ v }_{ F }^{ 2 } {k}_{y}^{ 2 } + {\left| \Delta ({ k }_{ z }) \right|}^2} \right) ^{ 2 } 
\end{align}
which has an analogue condition to eqn. (\ref{eq:cond1}) for nodal behavior:
\begin{equation} \label{eq:cond2}
 ({ \Delta  }_{ S }-{ \Delta  }_{ D })^{ 2 }<{ m }^{ 2 } - { v }_{ F }^{ 2 } {k}_{y}^{ 2 }<({ \Delta  }_{ S }+{ \Delta  }_{ D })^{ 2 } .
\end{equation}
The new feature of such a geometry is that, instead of point nodes, this architecture supports line nodes. For our particular choice of magnetization, the zeros lie in the $k_{z}-k_{y}$ plane.
The band for which this occurs is $\varepsilon_{-}$.  The resulting surfaces $ \pm \varepsilon_{-}(k_z,k_y) $ touch along a \emph{curve}, called a ``line-node", given by 
\begin{equation} \label{eq:curve}
{ v }_{ F }^{ 2 } {k}_{y}^{ 2 } + 2\Delta_S\Delta_D\cos(k_z d) = m^2-\Delta_S^2-\Delta_D^2  .
\end{equation}

The curve is always bounded in the $k_y$-direction,
$$   m^2 - ({ \Delta  }_{ S }+{ \Delta  }_{ D })^{ 2 } < { v }_{ F }^{ 2 } {k}_{y}^{ 2 } < m^2 - ({ \Delta  }_{ S }-{ \Delta  }_{ D })^{ 2 } .$$
Since this relation potentially places a minimum on $k_y$, the curve is not necessarily closed. The upshot is that the variation of magnetization leads to an evolution of the nodal line from being closed within a Brillouin zone to being open. Thus the low energy sector of such an architecture is highly tunable. We explore the properties, such as density of states, conductivity and magneto-oscillations, in the rest of the paper.

\subsection{Closed Line-Node}

\begin{figure}[ht]
\includegraphics[width= 0.8 \columnwidth]{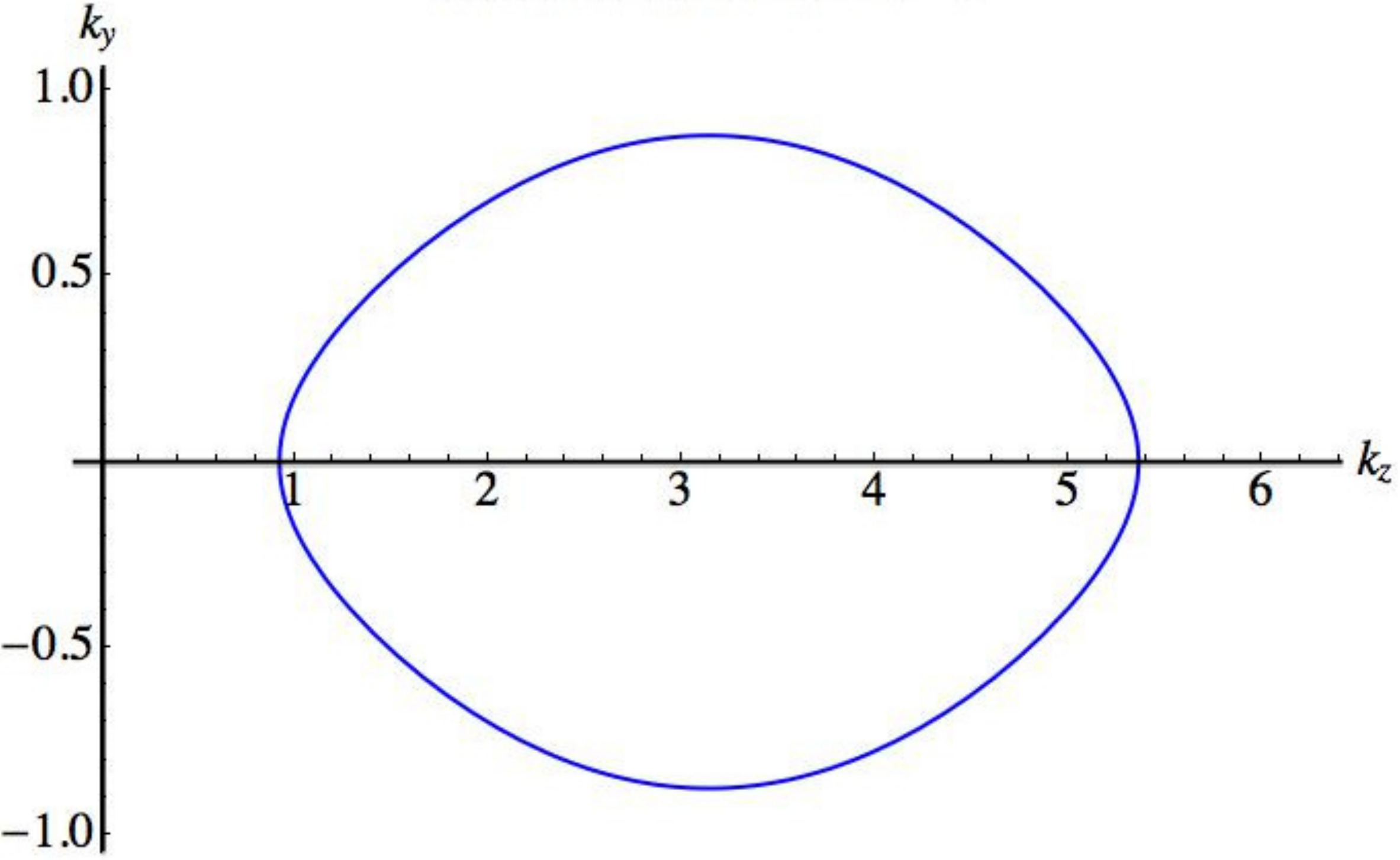}
\caption{An example of the nodal curve for the closed region of parameters.  The parameters chosen are $m=0.9, \Delta_S=0.6, \Delta_D=0.4$ (with $d=v_F=1$). } 
\label{fig:curveplot}
\end{figure}

\begin{figure}[ht]
\includegraphics[width= 0.8 \columnwidth]{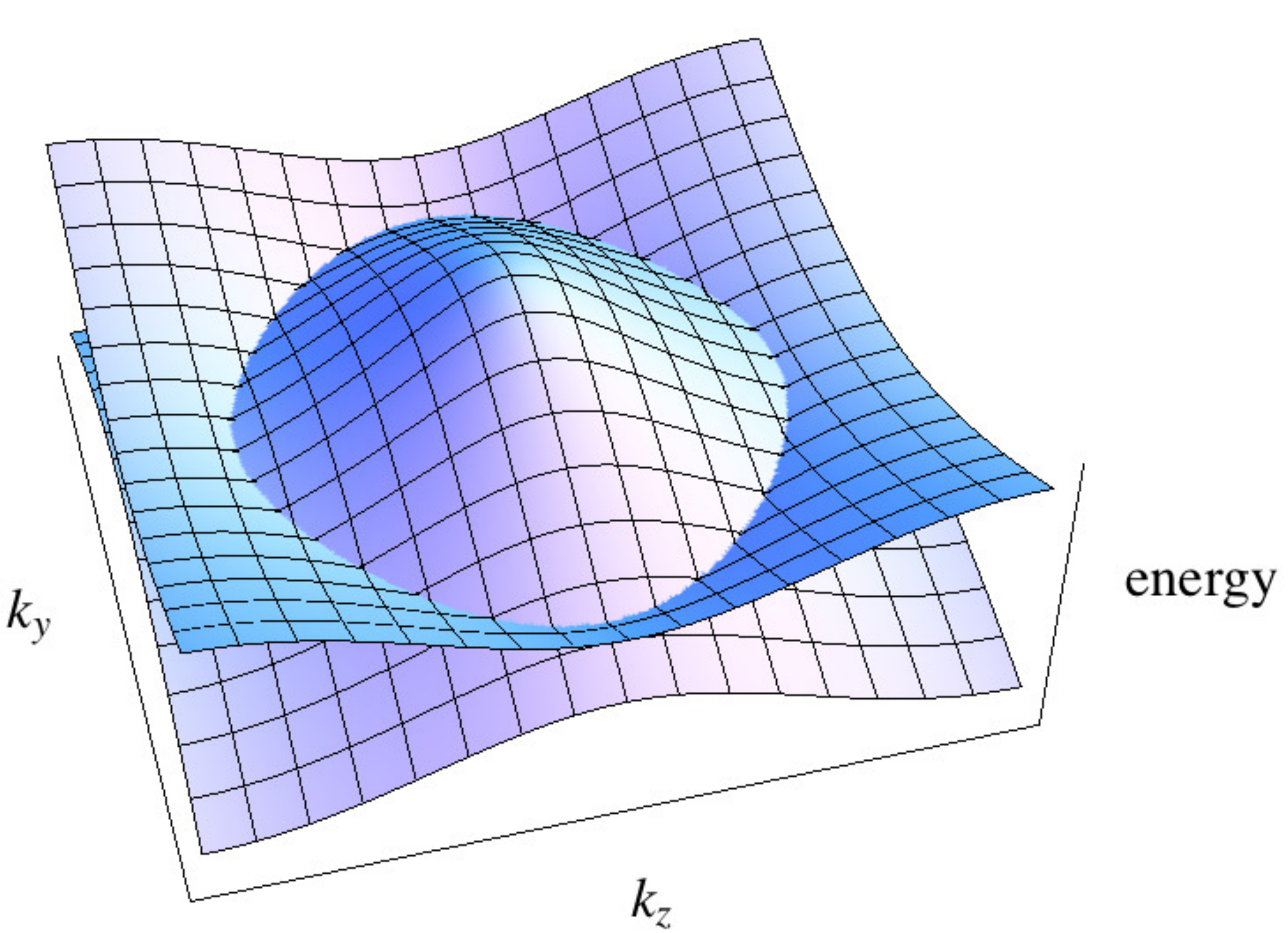}
\caption{The dispersion surfaces $\pm\varepsilon_{-}(k_z,k_y)$, showing the nodal curve where the top and bottom surfaces touch along $\varepsilon_{-}=0$.  The parameter values are the same as for fig. \ref{fig:curveplot}.}  
\label{fig:dsurf}
\end{figure}

Let us first examine the parameter space  $ |{ \Delta  }_{ S }-{ \Delta  }_{ D }| < m < { \Delta  }_{ S }+{ \Delta  }_{ D } $.  A characteristic nodal line is shown in fig. \ref{fig:curveplot}. To further examine the nature of the dispersion, we plot the energy as a function of $k_{z}-k_{y}$ for $k_{x}=0$ in fig. \ref{fig:dsurf}.

An interesting feature is that the dispersion is linear in momentum for deviations normal to the nodal curve. Parametrizing the curve as $(k_{z}^{0}, k_{y}^{0})$, which satisfy eqn. (\ref{eq:curve}), the dispersion as a function of deviation normal to the curve is

\begin{equation}    \label{eq:linDisp}
\varepsilon^{2}_{-}\approx v_{F}^{2}\delta k_{x}^{2} + \left({v_{F} k_{y}^{0}\over {m\cos(\theta_{0})}}\right)^{2} v_{F}^2 \delta k_{\perp}^{2}
\end{equation}
where $\tan(\theta_{0}) = \Delta_{D}\Delta_{S}d\sin(k_{z}^{0}d)/v_{f}^{2}k_{y}^{0}$.

\begin{figure}[ht]
\includegraphics[width=\linewidth]{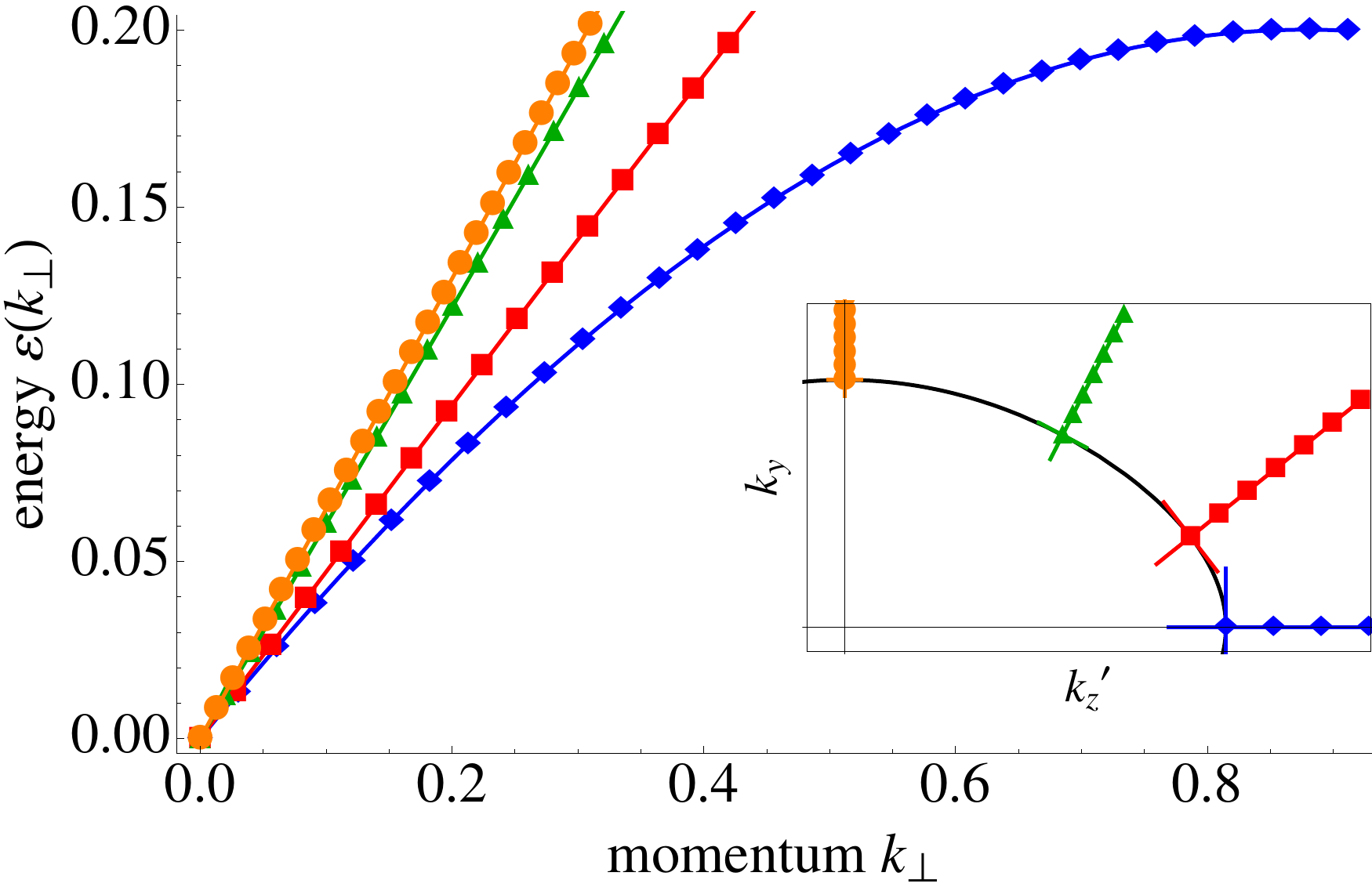}
\caption{The variation of energy as a function of momentum perpendicular to the nodal line at four representative points is shown. Parameters remain as in fig. \ref{fig:curveplot}.}
\label{fig:linlife}
\end{figure}

The energy scale at which the deviation from linearity becomes substantial is also a function of where one is on the nodal curve. Thus an effective linear dispersion is valid only in an energy window which is the minimum of this function. To display this variation we plot the dispersion along various cuts across the nodal line in fig.\ref{fig:linlife}. We occasionally use $k_{z}' \equiv k_{z}-{\pi \over d}$ as a convenient variable. The ratio of the velocities along the cuts for $k_{y}=0$ and $k_{z}'=0$ is

\begin{equation}
{v_{\perp}(k_{y}=0)\over{ v_{\perp}(k_{z}'=0)}} = {md\over {2v_{F}}}\sqrt{\left({\Delta_{S}+\Delta_{D}\over m}\right)^{2}-1} \thinspace.
\end{equation}
This is always less than one implying that the corrections to linearity are more pronounced along the growth axis of the heterostructure. Furthermore the monotonic increase of velocity from $k_{z}$ to $k_{y}$ axis implies that the density of states also varies along the node.

\begin{figure}[ht]
\includegraphics[width=\linewidth]{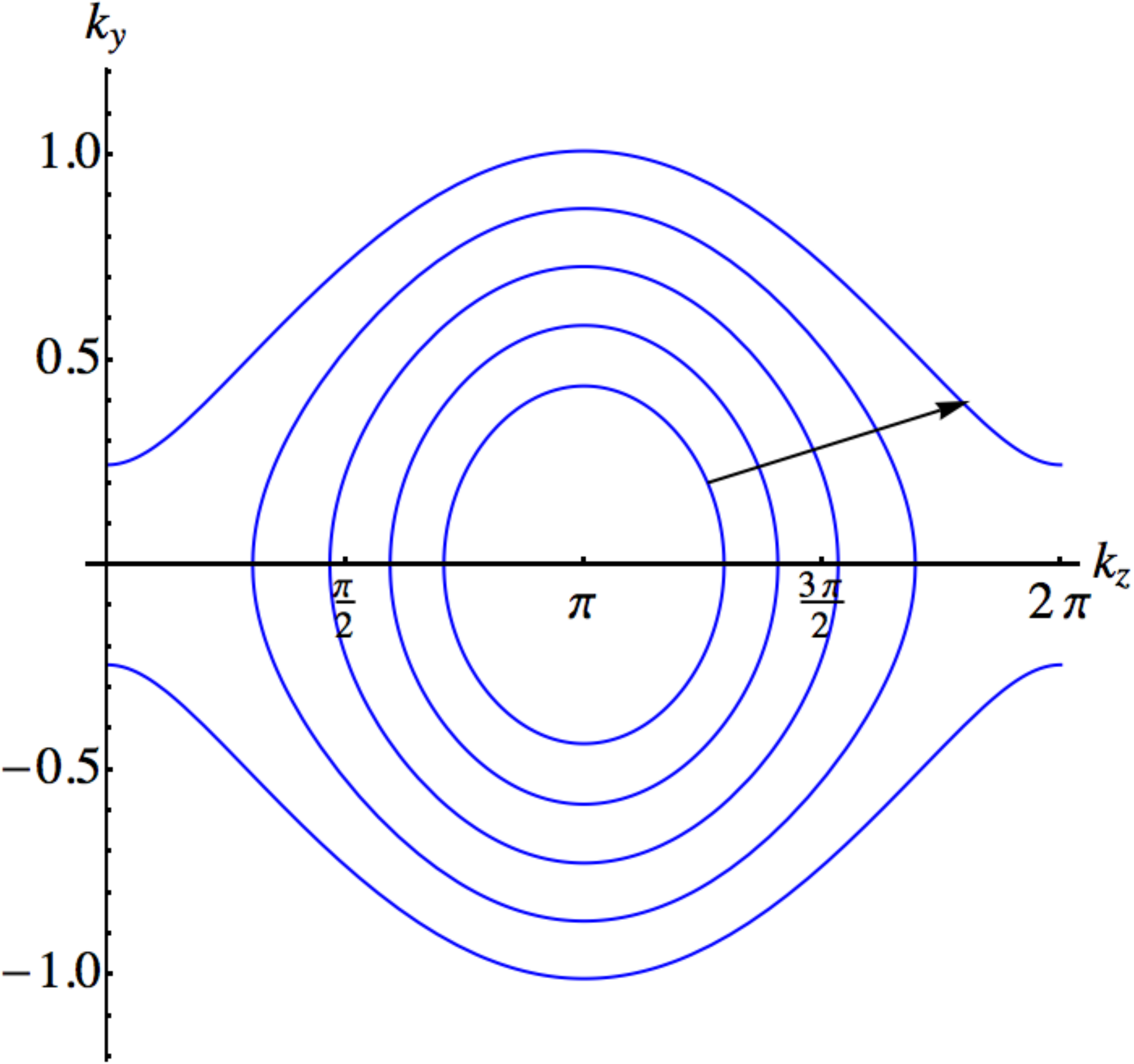}
\caption{The shape of the closed line-node in fig. \ref{fig:curveplot} is plotted as the field parameter $m$ is changed, while keeping $\Delta_S$ and $\Delta_D$ fixed.  The arrow shows the direction of increasing $m$. The closed curves have $m< \Delta_S+\Delta_D$ while the open curve violates this condition.}
   \label{fig:manip-node}
\end{figure}

For a given device $\Delta_{S}$ and $\Delta_{D}$ are hard to tune but the magnetization can be modified. As a consequence the nodal structure evolves and as the magnetization is increased the curve extends out along $k_{z}$ until it reaches the edge of the Brillouin zone. This is shown in fig. \ref{fig:manip-node}. In general, analytic solutions are not possible. Under certain approximations the density of states and conductivity can be obtained in closed form. These are

\begin{enumerate}
\item The width of the nodal curve is ``small" : \\ $\cos (k_z d) \approx -1 + \frac{(k_z d - \pi)^2}{2}$  \label{num:asum1}
\item The parameter $m$ is ``large" : \\ $m \gg | \Delta_S - \Delta_D | \equiv \Delta \Rightarrow  \Delta / m \approx 0$ \label{num:asum2}
\end{enumerate}
Together, these greatly simplify the positive and negative dispersions that lead to the node, $\pm \varepsilon_{-}(\vec{k}) \equiv \pm \varepsilon (\vec{k})$ in eqn. (\ref{eq:dis2}).  From here, the density of states is found by taking the derivative $g(\varepsilon) = \frac { d }{ d\varepsilon  } N(\varepsilon )$, where an integral must be done:

\begin{equation}    \label{eq:Ndef}
N(\varepsilon ) = \int \frac{d^3 k}{(2 \pi )^3}   \Theta[ \varepsilon -  \varepsilon (\vec{k}) ].
\end{equation}

With the above approximations, this integral becomes the volume of a torus in momentum-space with major radius $m$ and minor radius $\varepsilon$, which can be calculated analytically to give a linear density of states (DoS): $g(\varepsilon) \propto \varepsilon$.  The constant of proportionality (the slope of the DoS) comes out to be linear with respect to $m$.  The resultant  DC conductivity (using the Kubo formula, after the Self-Consistent Born Approximation)  is also linear in $m$\cite{bhb}.

\begin{figure}[ht]
\begin{tabular}{l | r}
\includegraphics[width=0.25\textwidth]{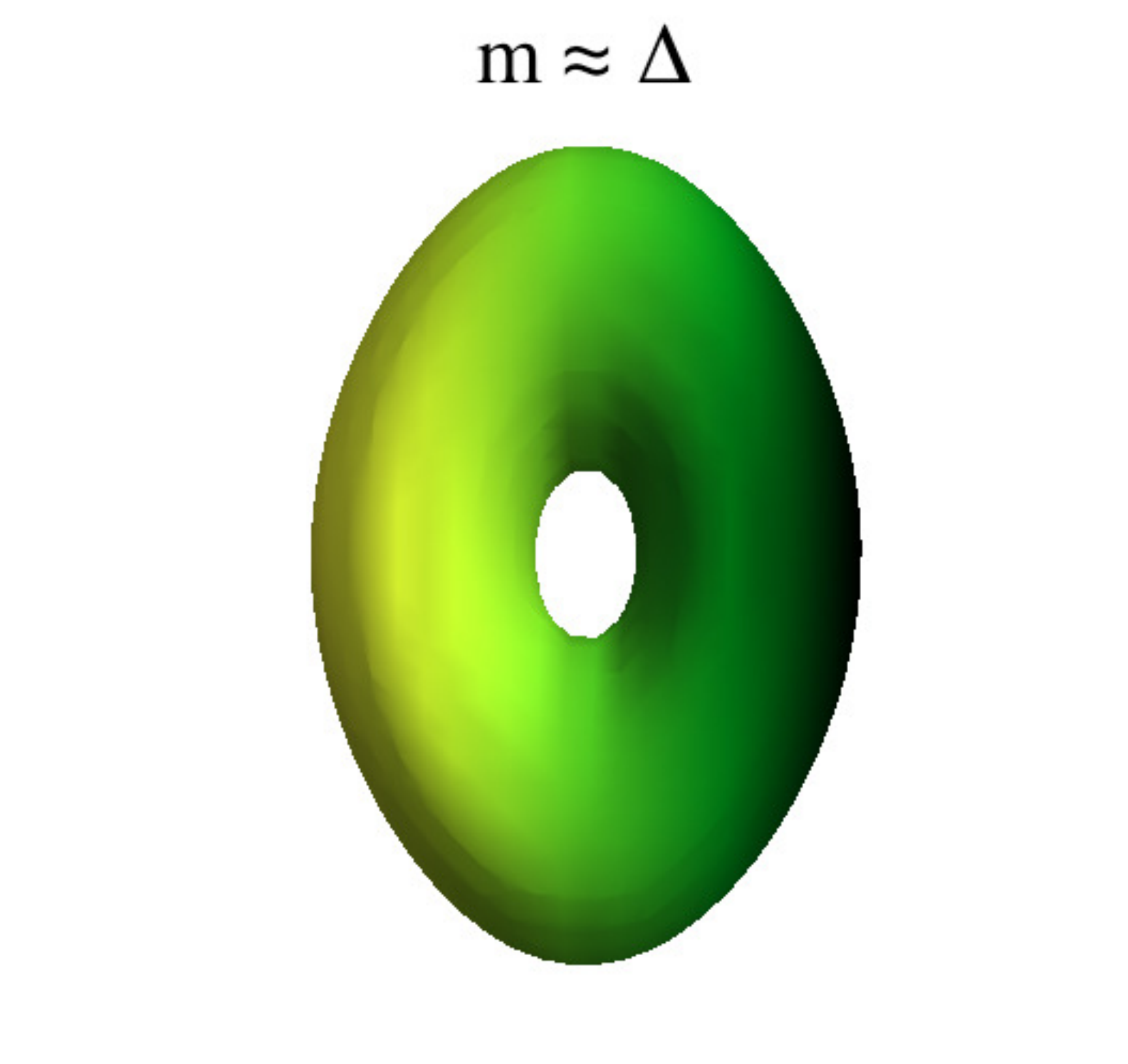} &  \includegraphics[width=0.25\textwidth]{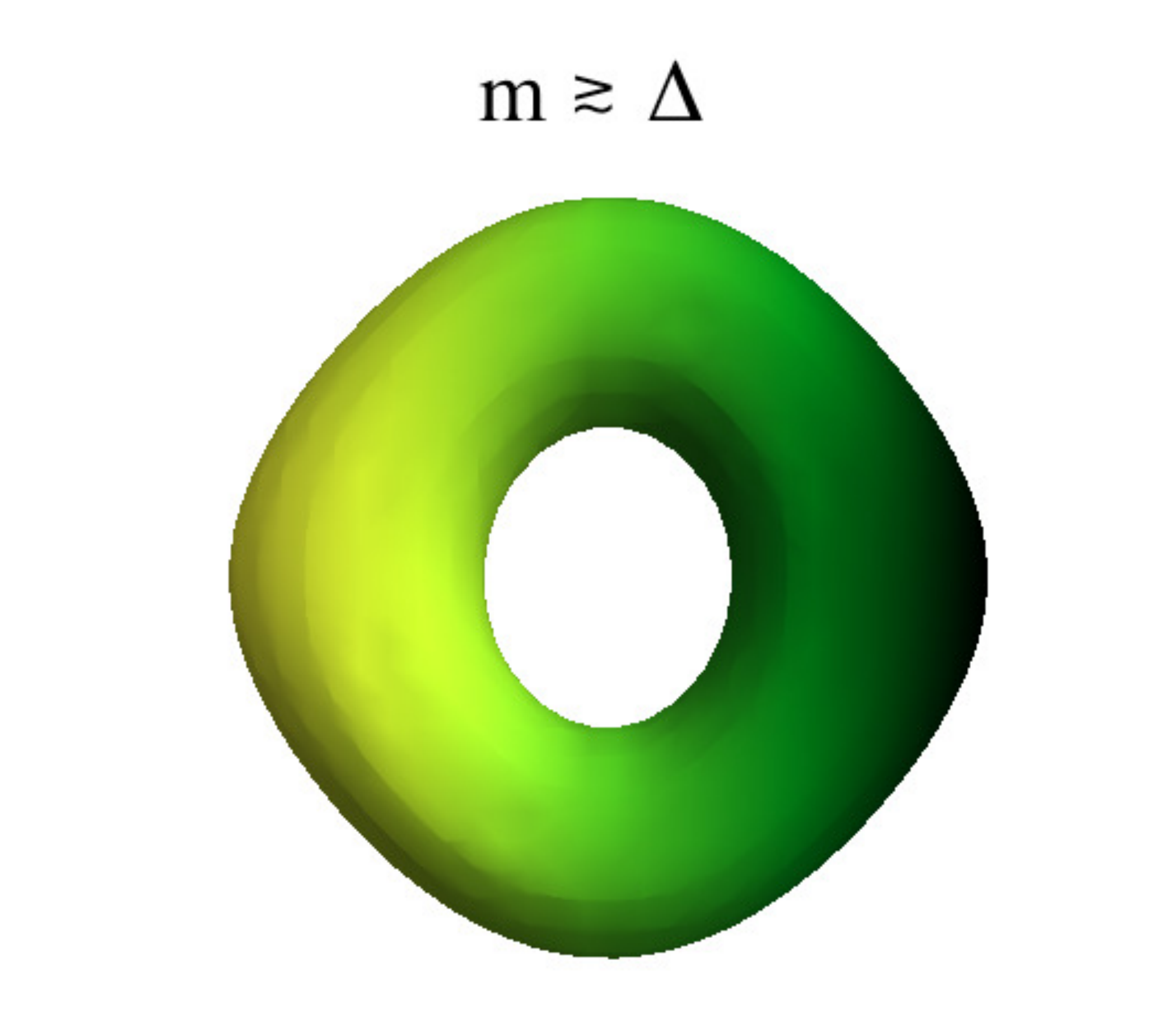}  \\
\hline
\includegraphics[width=0.25\textwidth]{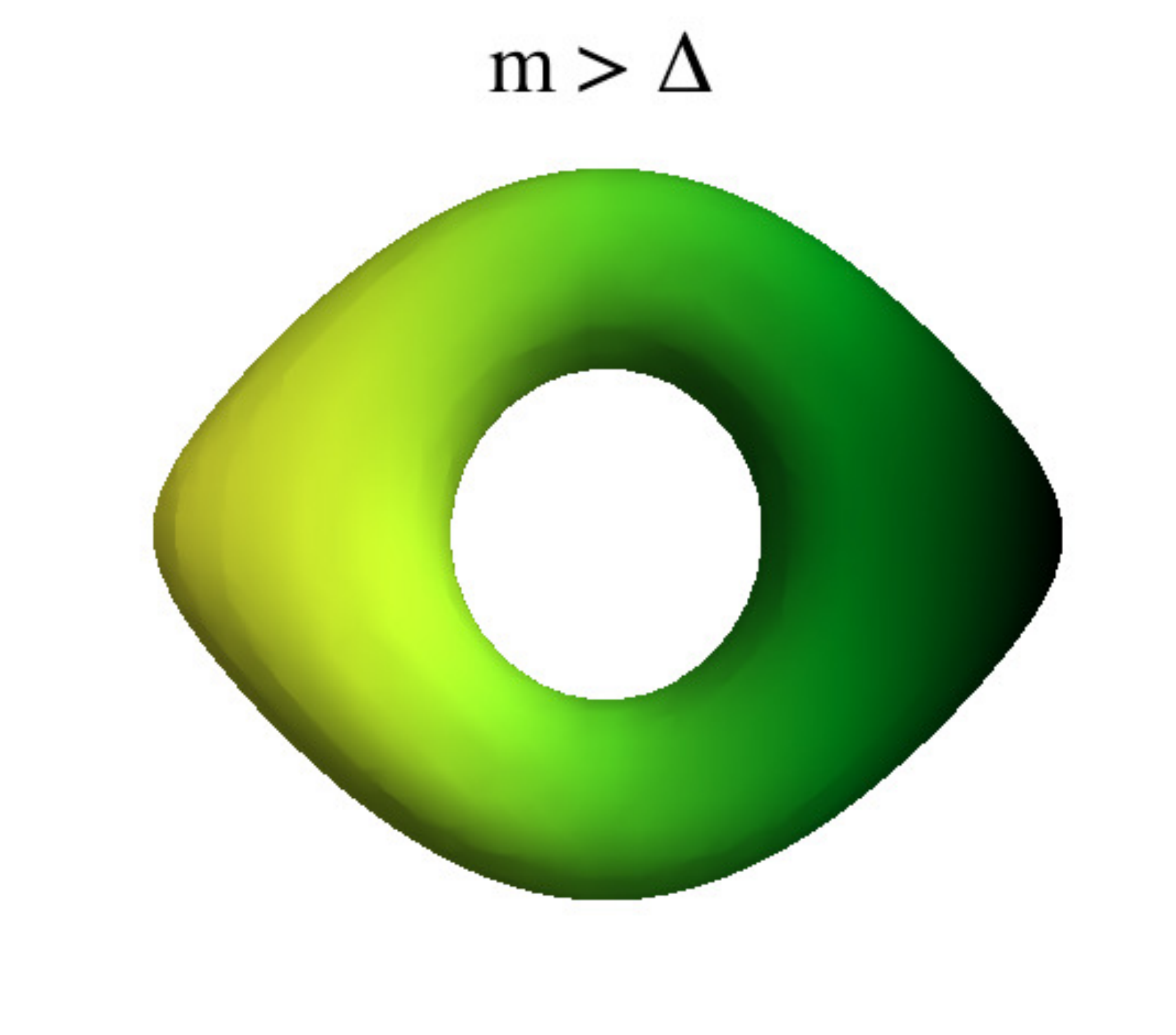} &  \includegraphics[width=0.25\textwidth]{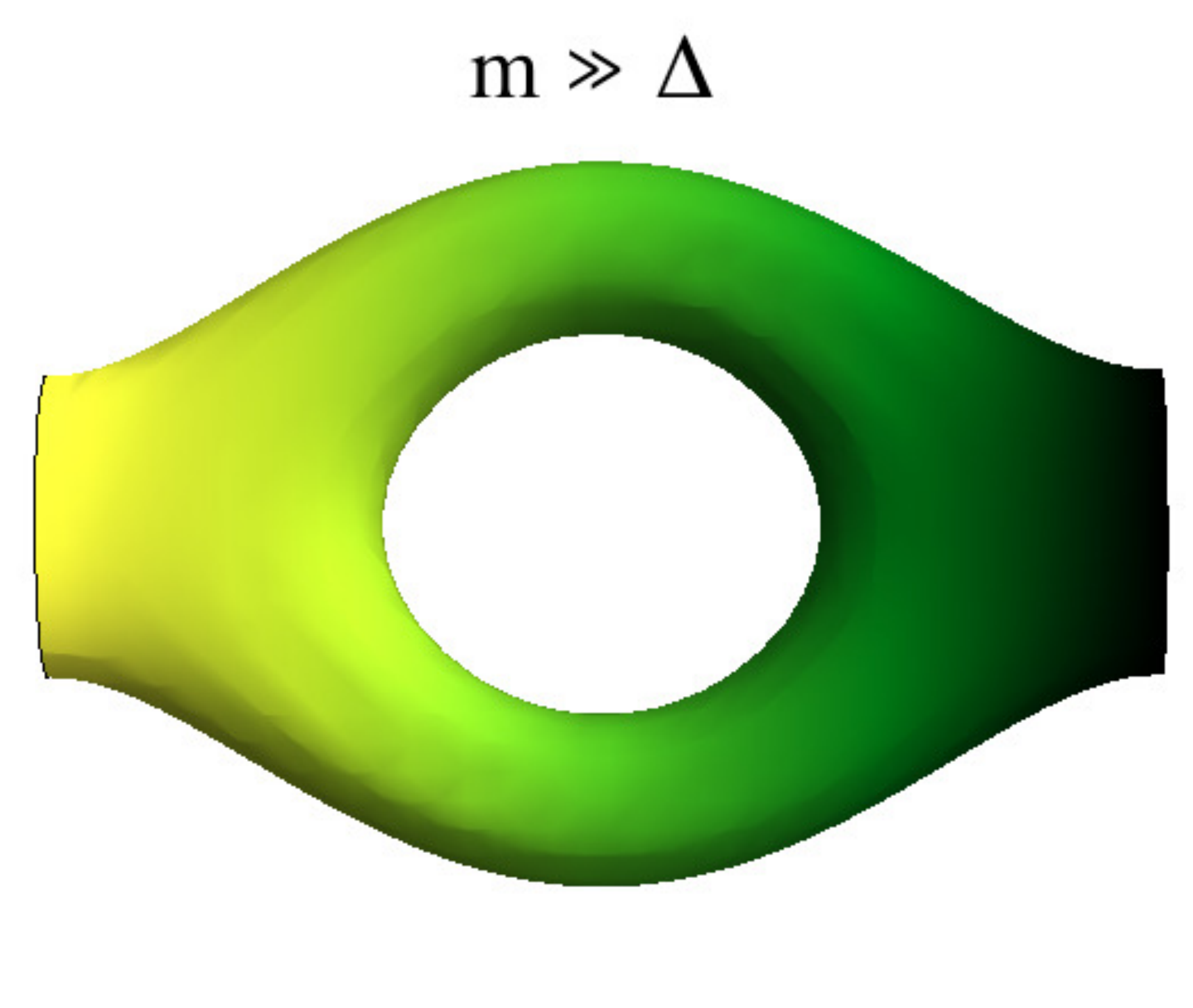}  \\
\end{tabular}

 \caption{ The volume enclosed by a surface of energy $\varepsilon$ in momentum space,  $N(\varepsilon)$, is plotted while varying the field parameter $m$. A perfect torus is obtained for a particular value of $m$.  For smaller $m$, the torus gets squeezed, while for larger values the torus gets stretched (along the $k_z$-axis). }
   \label{fig:manip-dos}
\end{figure}
The analytic expressions provide an interesting insight into the behavior of this device. The low temperature properties are all functions of the magnetization $m$. Thus tuning this parameter allows for the modification of transport and thermodynamic response.  To get an accurate description numerical methods need to be employed as the approximations stated above are valid only for a finite intermediate window of $m$. As shown in fig. \ref{fig:manip-dos}, the equal energy surfaces have significant deviations from a uniform torus.  Thus the DC conductivity will match the analytical expression for a finite range of magnetization. 

\begin{figure}

 \includegraphics[width=0.9\columnwidth]{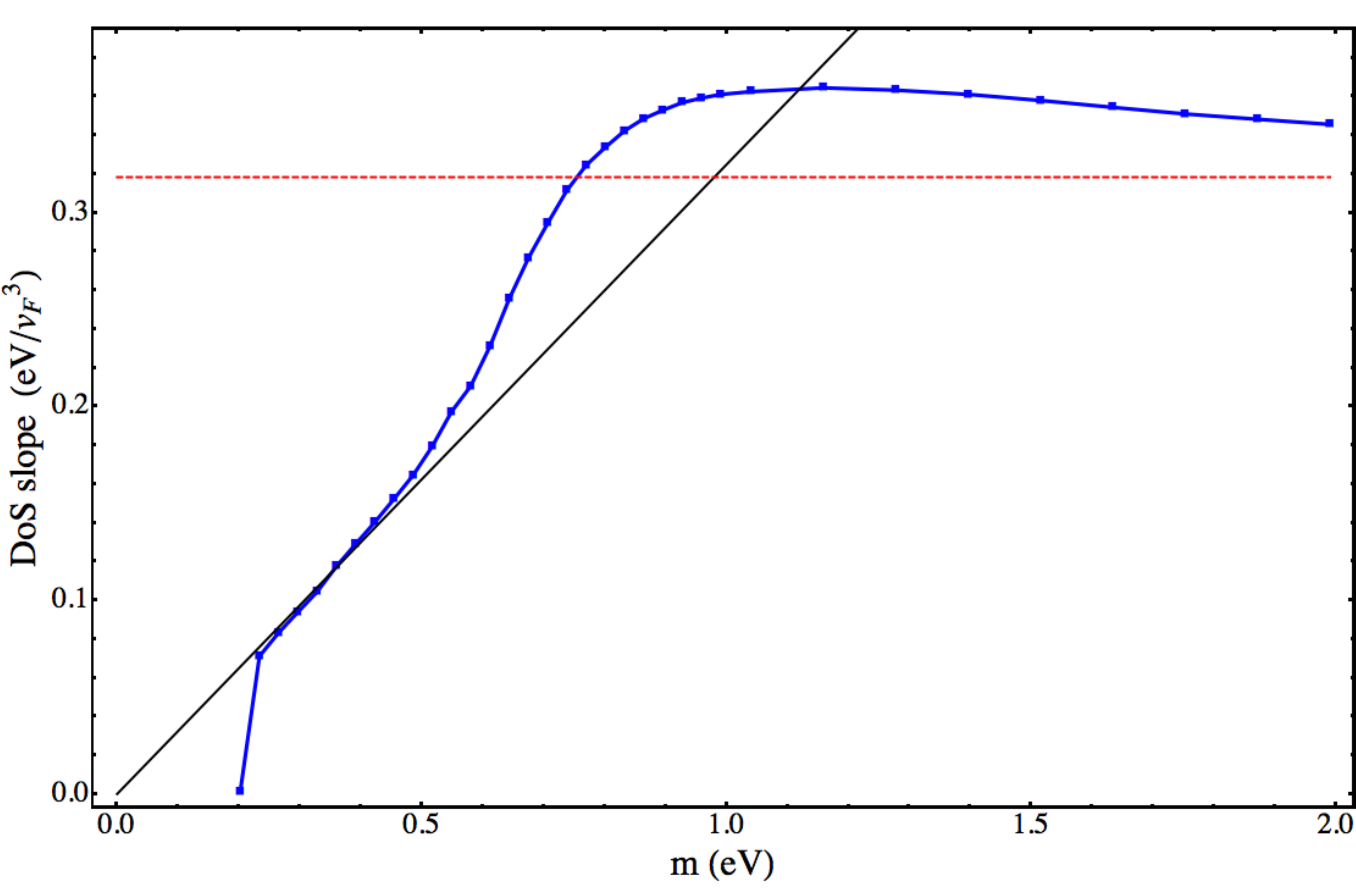}

\caption{
      The slope of the linear (low-energy) density of states is plotted as a function of $m$.  The points are obtained from numerical evaluation of the integral in eqn. (\ref{eq:Ndef}). The black line is the analytical result from [\onlinecite{bhb}], while the dashed line shows the asymptotic value. }
   \label{fig:slopeVm}
\end{figure}

To characterize the device better we employ numerical solutions for the full dispersion. Unless otherwise specified the parameters used are $\Delta_{S}=\SI{0.6}{\electronvolt}$ and $\Delta_{D}=\SI{0.4}{\electronvolt}$. The results are qualitatively identical for different choices of parameters. We first compare the slope of the density of states near $\varepsilon=0$ to that obtained analytically. This is shown in fig. \ref{fig:slopeVm}.  A monotonically increasing slope is obtained as long as the nodal line remains closed. Once the nodal line hits the Brillouin zone boundary, the slope is roughly constant. We discuss the open node case in the next section.

\begin{figure}[ht]
\includegraphics[width=1.05\linewidth]{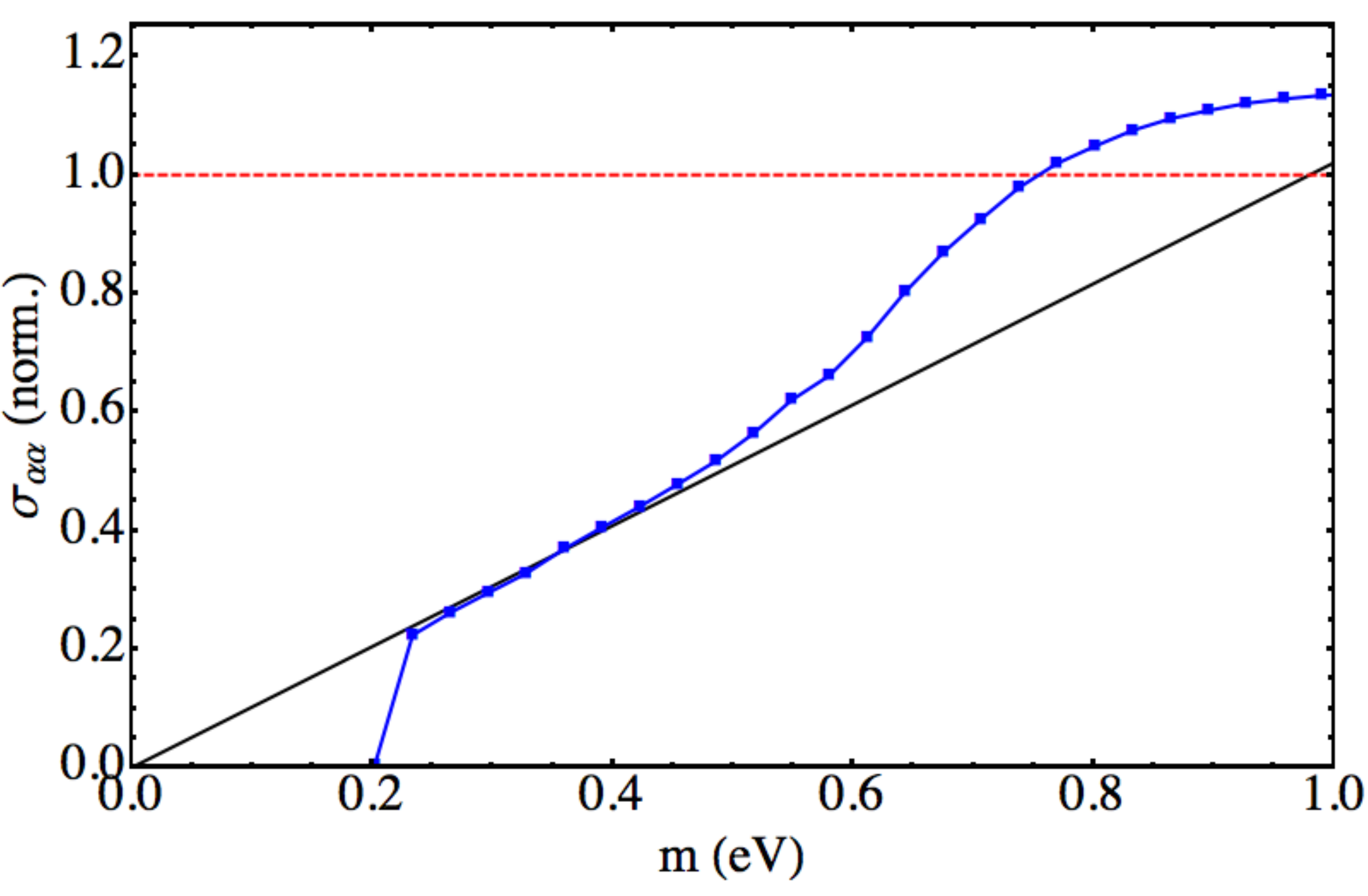}
\caption{For large values of $m$, where the nodal line is open, the conductivity  is roughly constant.  Here we plot the conductivity divided by the asymptotic value as a function of $m$. These normalized curves are isotropic but the asymptotic values themselves are different for parallel and perpendicular directions with respect to the growth axis of the multilayer.}
\label{fig:polyCond}
\end{figure}

For point-like impurities within the Self-Consistent Born approximation using Kubo formalism, the conductivity is proportional to the slope of the density of states. The linear DoS is found numerically over the entire range of $m$ using eqn. (\ref{eq:Ndef}), resulting in the conductivity shown in fig. \ref{fig:polyCond}.  For large values of $m$, the constant slope given by eqn. (\ref{eq:dosApx}) leads to a conductivity that's also independent of $m$. This asymptotic value depends on the direction as there is an anisotropy in velocity parallel versus perpendicular to the growth axis of the device. To display the generic behavior we plot the ratio of the conductivity to the asymptotic value as a function of $m$. 

The asymptotic values are obtained from the linear density of states with a constant slope (for details see the next section) and given by

\begin{equation}    \label{eq:asympCond}
\sigma_{\alpha \alpha}={2 e^2 \over h} {v_{F,\alpha}^2 \over \pi d v_F^2}
\end{equation}

where $v_{F,x}=v_{F,y}=v_F$ and $v_{F,z}=d \sqrt{\Delta_S \Delta_D}$.

\begin{figure}[ht]
\begin{tabular}{l  r}
\includegraphics[width=0.31\textwidth]{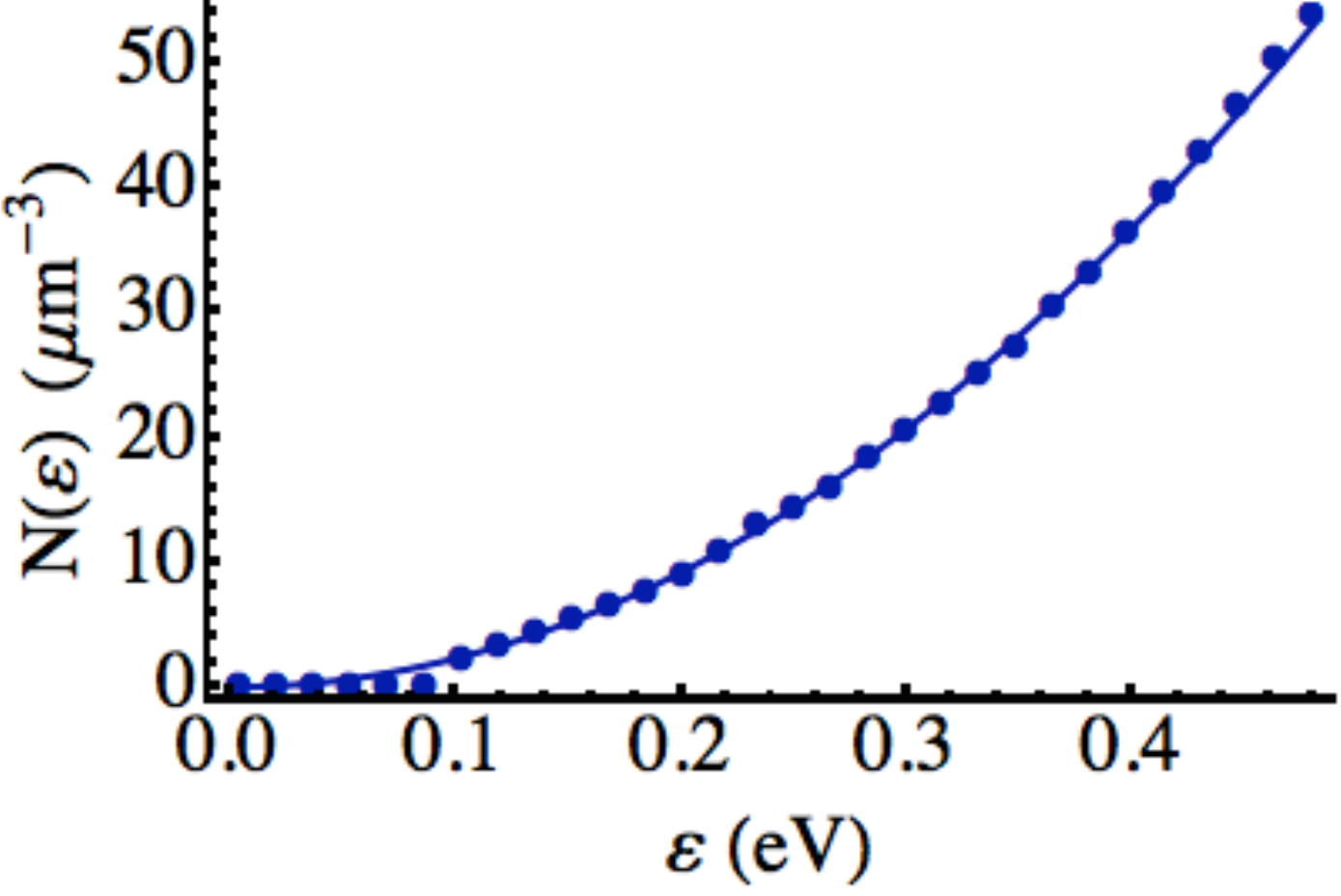}  \\
\includegraphics[width=0.31\textwidth]{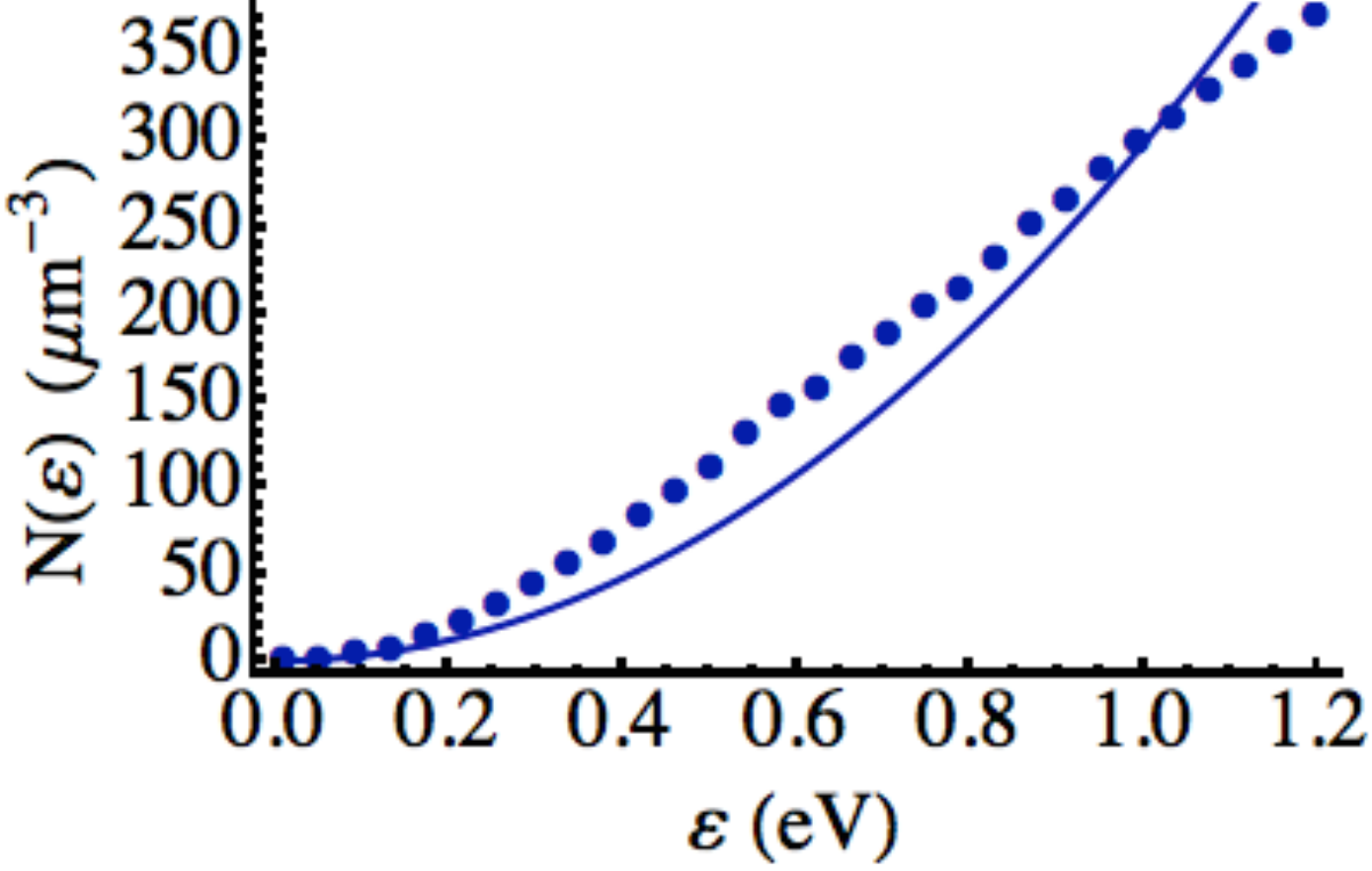} 
\end{tabular}

 \caption{
      We plot $N(\varepsilon)$ for two different values of $m$. The top figure is for the case where we have a closed nodal line where the quadratic dependence is evident reflecting a linear density of states. For large $m$, for open line nodes, the quadratic dependence crosses over to a linear behavior very quickly as one departs from the node. This means that a linear DoS is trustworthy only for very low energies when $m \gg  { \Delta  }_{ S }+{ \Delta  }_{ D }$ (see fig. \ref{fig:manip-open}).
      }
   \label{fig:nplots}
\end{figure}

To better understand the nature of the density of states we also examine $N(\varepsilon)$ itself (see eqn. (\ref{eq:Ndef})) for various values of $m$. Its quadratic dependence of $\varepsilon$ yields a linear density of states. As noted in the discussion of the dispersion and shown in fig. \ref{fig:linlife}, the deviation from linear dispersion occurs for small distances away from the node. The change in the dispersion is also evident in the evolution of $N(\varepsilon)$ plotted in fig. \ref{fig:nplots}. For small $m$ a quadratic behavior is seen, but becomes linear as $m$ is increased. The change in the density of states is reflected in the conductivity.

\subsection{Open Line-Node}

We now turn to the regime of large $m$. This section examines the case $m> { \Delta  }_{ S }+{ \Delta  }_{ D }$. For these values of $m$ the node touches the sides of the first BZ along the $k_z$-axis. This implies that the analytic result, $\text{slope} \propto m$, of the previous section does not apply but the numerical techniques can be used. The density of states at low energies remains linear, but its slope is constant as seen in fig. \ref{fig:slopeVm}. To evaluate the constant we use eqns. (\ref{eq:linDisp}) and (\ref{eq:curve}) assuming large $m$.  Adding $2 \Delta_S \Delta_D$ to both sides of eqn. (\ref{eq:curve}) and with $m \gg | \Delta_S - \Delta_D |$, we drop $(\Delta_S - \Delta_D)^2$ from the right hand side of eqn. (\ref{eq:curve}) to get 
\begin{equation}    \label{eq:curveApx}
({ v }_{ F } {k}_{y}^0)^{ 2 } + 2\Delta_S\Delta_D \left( 1+\cos(k_z^0 d) \right) \approx m^2
\end{equation}

where the notation $k_{\alpha}^0$ refers to the points on the nodal line. The second term in eqn. (\ref{eq:curveApx}) is bounded by the value of the cosine, and is small when  $m^2 \gg 4 \Delta_S \Delta_D$. Therefore $ {v }_{ F } {k}_{y}^0 \approx \pm m$. In other words, the open nodes evolve into straight lines stretching across the Brillouin zone at a fixed value of $k_{y}$. Similarly $\tan{\theta_0}= \Delta_{D}\Delta_{S}d\sin(k_{z}^{0}d)/v_{f}^{2}k_{y}^{0}$ is bounded by the sine function and for $m \gg {\Delta_S \Delta_D d / v_F}$, $\theta_0 \ll 1$. Asymptotically  eqn. (\ref{eq:linDisp})  simplifies to

\begin{equation}    \label{eq:linDispApx}
\varepsilon^{2}_{-}\approx v_{F}^{2}\delta k_{x}^{2} + v_{F}^2 \delta k_{y}^{2}
\end{equation}

which, remarkably, is identical to the graphene dispersion. Setting $q=v_F \sqrt{\delta k_{x}^{2}+\delta k_{y}^{2}}$, the DoS is

\begin{eqnarray}    \label{eq:dosApx}
g(\varepsilon) &=& 2 \int_0^{2 \pi / d} {dk_z \over 2\pi} \int_0^{2 \pi } {d\phi \over 2\pi} \int {q dq \over 2\pi v_F^2} \delta(\varepsilon - q)  \nonumber   \\
    &=&{ \varepsilon \over \pi d v_F^2 }.
\end{eqnarray}
The resulting conductivity reflects this behavior and is nearly independent of $m$ for $m>\SI{1}{\electronvolt}= \Delta_{S} + \Delta_{D}$. The conductivity varies appreciably only when $ |{ \Delta  }_{ S }-{ \Delta  }_{ D }| < m < { \Delta  }_{ S }+{ \Delta  }_{ D } $. An interesting aspect of this device is its sensitivity to changes in magnetization. The change in conductivity from zero to the maximum value given in eq. (\ref{eq:asympCond}) occurs over the change in exchange splitting of $2\Delta_{S}$ or $2\Delta_{D}$, whichever is smaller. This sensitivity is a generic feature of such nodal semimetals. A minimum time reversal breaking field is needed to close the gap and further increase leads to the nodal line spanning the Brillouin Zone. Over this energy window the conductivity changes from zero to the asymptotic value.

\begin{figure}[ht]
\begin{tabular}{l | r}
\includegraphics[width=0.25\textwidth]{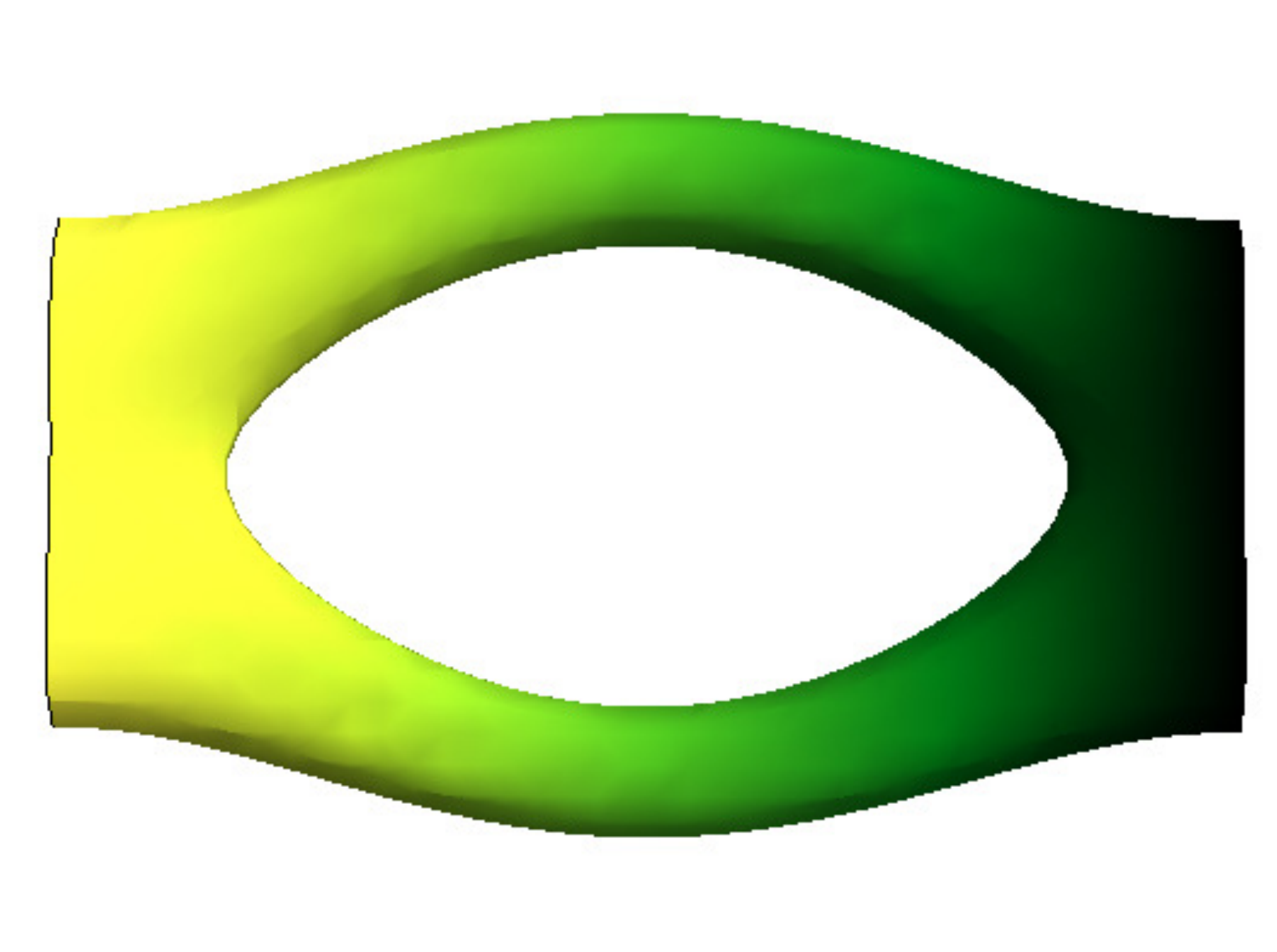}  &  \includegraphics[width=0.25\textwidth]{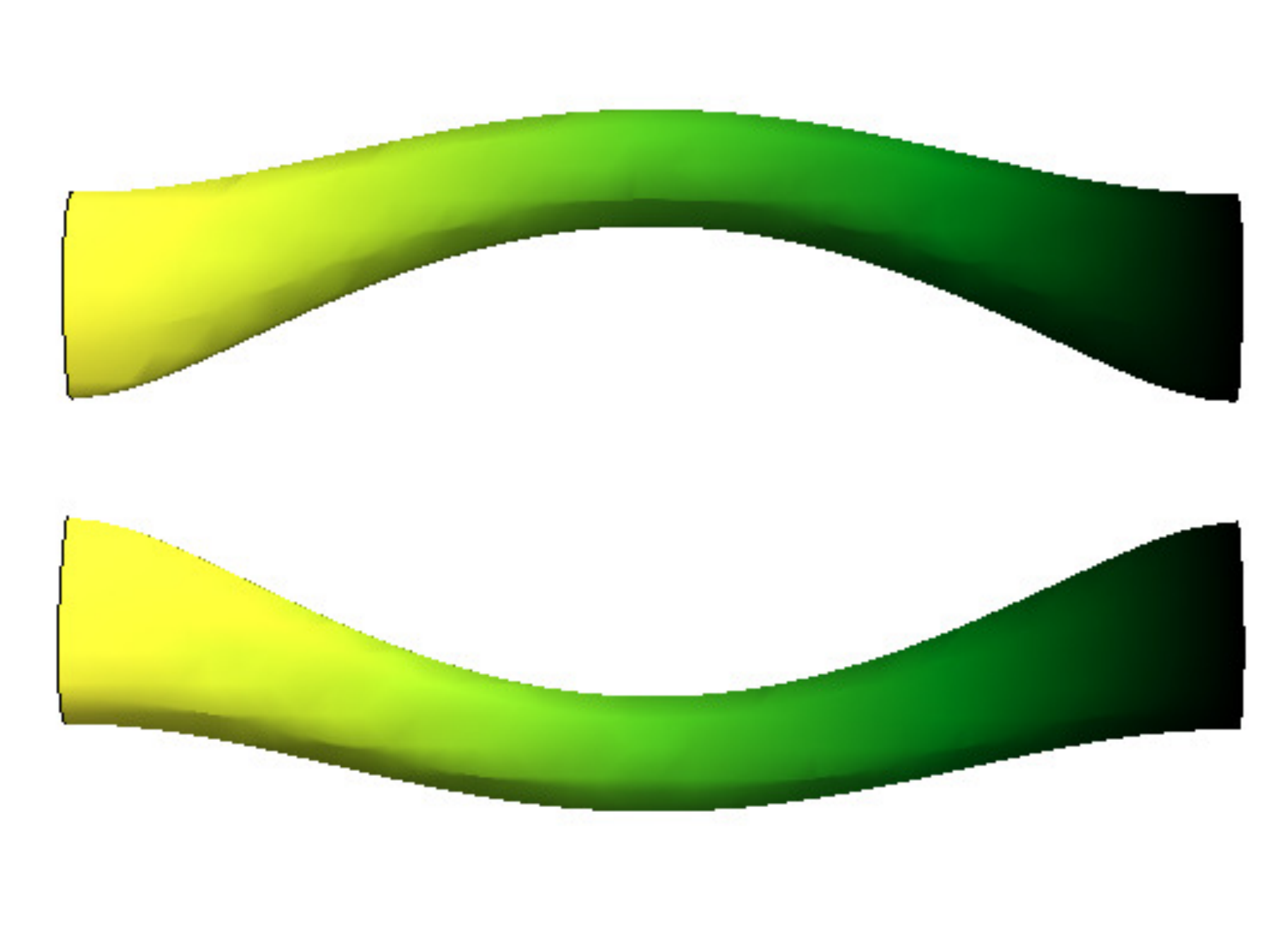}  \\
\end{tabular}
\caption{Constant energy surfaces for the same parameters as in fig. \ref{fig:slopeVm}, but with $m=\SI{1.15}{\electronvolt}>\Delta_{S} + \Delta_{D}$, and a larger $m=\SI{1.28}{\electronvolt} \gg \Delta_{S} + \Delta_{D}$. As $m$ increases the topology changes to disconnected surfaces.}

\label{fig:manip-open}
\end{figure}

The evolution of the fermi surface as seen in fig. \ref{fig:manip-dos} is measurable in magnetic oscillation experiments. Before we turn to the discussion of the expected behavior, it is worthwhile to note the shape of the equal energy surface for even larger $m$. In fig. \ref{fig:manip-open} we see that the the volume enclosed gets disconnected on increasing $m$. Thus the fermi surface of the doped system goes from a closed torus to two disconnected tubes, with an intermediate state where the outer surface of the torus is not closed within a Brillouin zone. Implications of these on quantum oscillations are discussed in the next section.

\subsection{Quantum Oscillations} \label{sec:dehaas}

The evolution of the low energy sector  as a function of the magnetization can be probed for systems with finite carrier density. For closed nodal lines the fermi surface
has the topology of a torus, whose axis is parallel to the direction of magnetization. In the presence of an external magnetic field, the density of states is oscillatory. For large densities (i.e. Landau level index corresponding to Fermi energy is large), the oscillation is periodic in 1/B with a frequency proportional to the extremal area $A_{e}$ of the fermi surface perpendicular to the applied field:
\begin{equation}   \label{eq:magFreq}
f_{\frac{1}{B}} = \frac{\hbar c}{2 \pi e} A_{e}(\varepsilon_F).
\end{equation}

\begin{figure}[ht]
\includegraphics[width=1.05\linewidth]{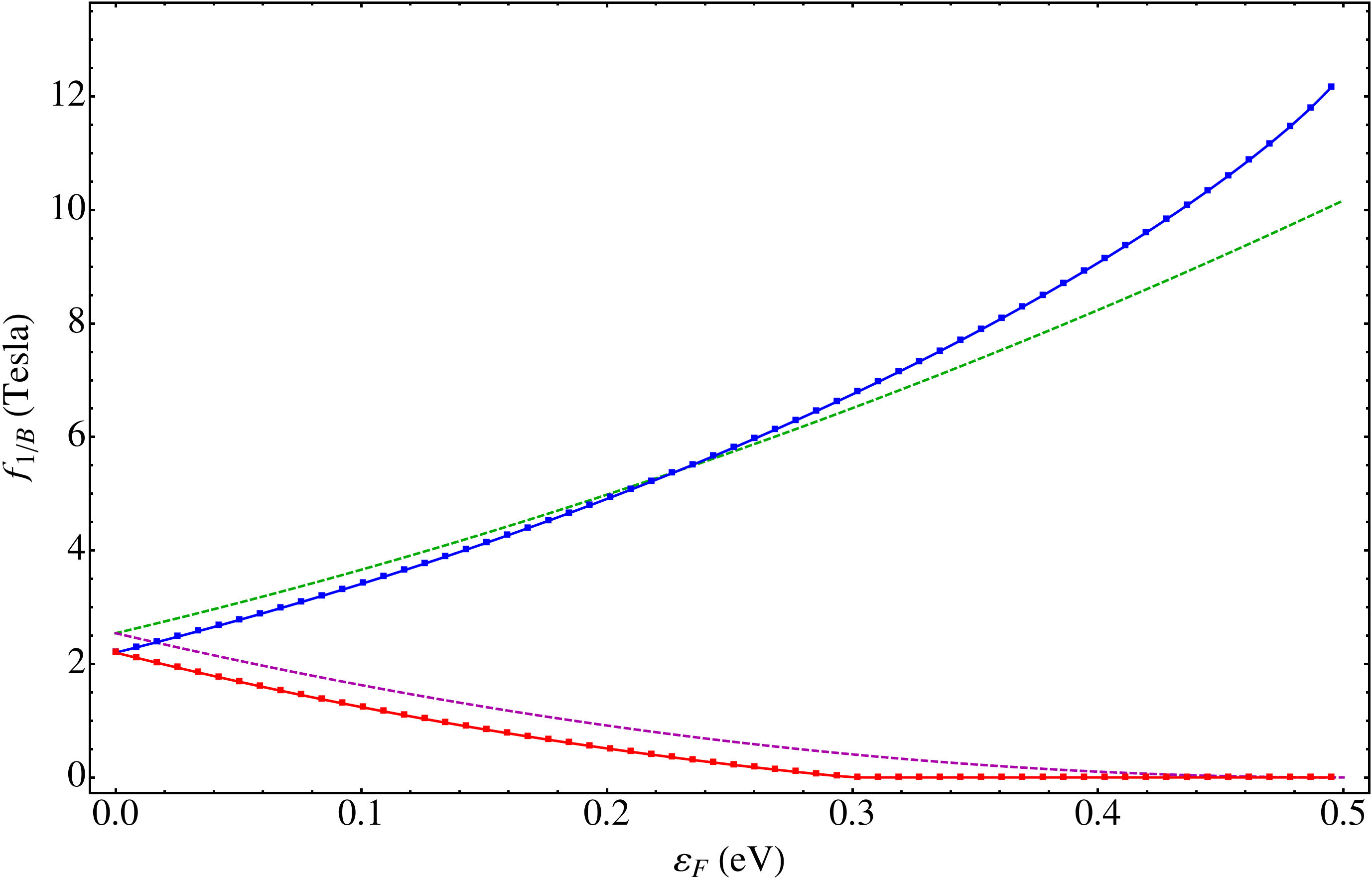}
\caption{Frequency for quantum oscillations for magnetic fields applied along the $x$-direction is plotted as a function of energy. The top curve shows the result corresponding to the maximal area, while the bottom corresponds to the minimal area. Dashed curves are obtained in the approximation of treating the surfaces as a circular torus. The parameters are the same as above with $m=\SI{0.5}{\electronvolt}$, taking $v_{F}=\SI{e4}{\meter\per\second}$ and $d=\SI{100}{\nano\meter}$. }

\label{fig:planarFreq}
\end{figure}

We consider two cases motivated by the geometry of equal energy surfaces. For magnetic field along the direction of the magnetization $m$ ($x$-axis in our example), there are two frequencies at small magnetization while for large magnetization one of two contributing orbits changes from a closed to an open one. In fig. \ref{fig:planarFreq} the two frequencies are plotted as a function of fermi energy. The fermi surface has the topology of a torus. For a field along the axis there are two extremal areas corresponding to the inner and outer circles. As the fermi energy increases the inner circle shrinks while the outer one grows. The smaller frequency vanishes when $\varepsilon_F = m-|\Delta_{S}-\Delta_{D}|$.

Alternatively, a field could be applied in the $z$-direction instead (i.e. along the growth direction). There is only one extremal orbit in this case. More precisely, there are two identical areas that contribute the same extremal area.  As one increases the doping or carrier concentration, these areas grow approaching one another. At a critical value of the fermi energy the two merge, and the resulting orbit continues to be extremal. Thus the frequncy doubles at the critical value $\varepsilon_F = m-|\Delta_{S}-\Delta_{D}|$.

\begin{figure}
\includegraphics[width=1.02 \linewidth]{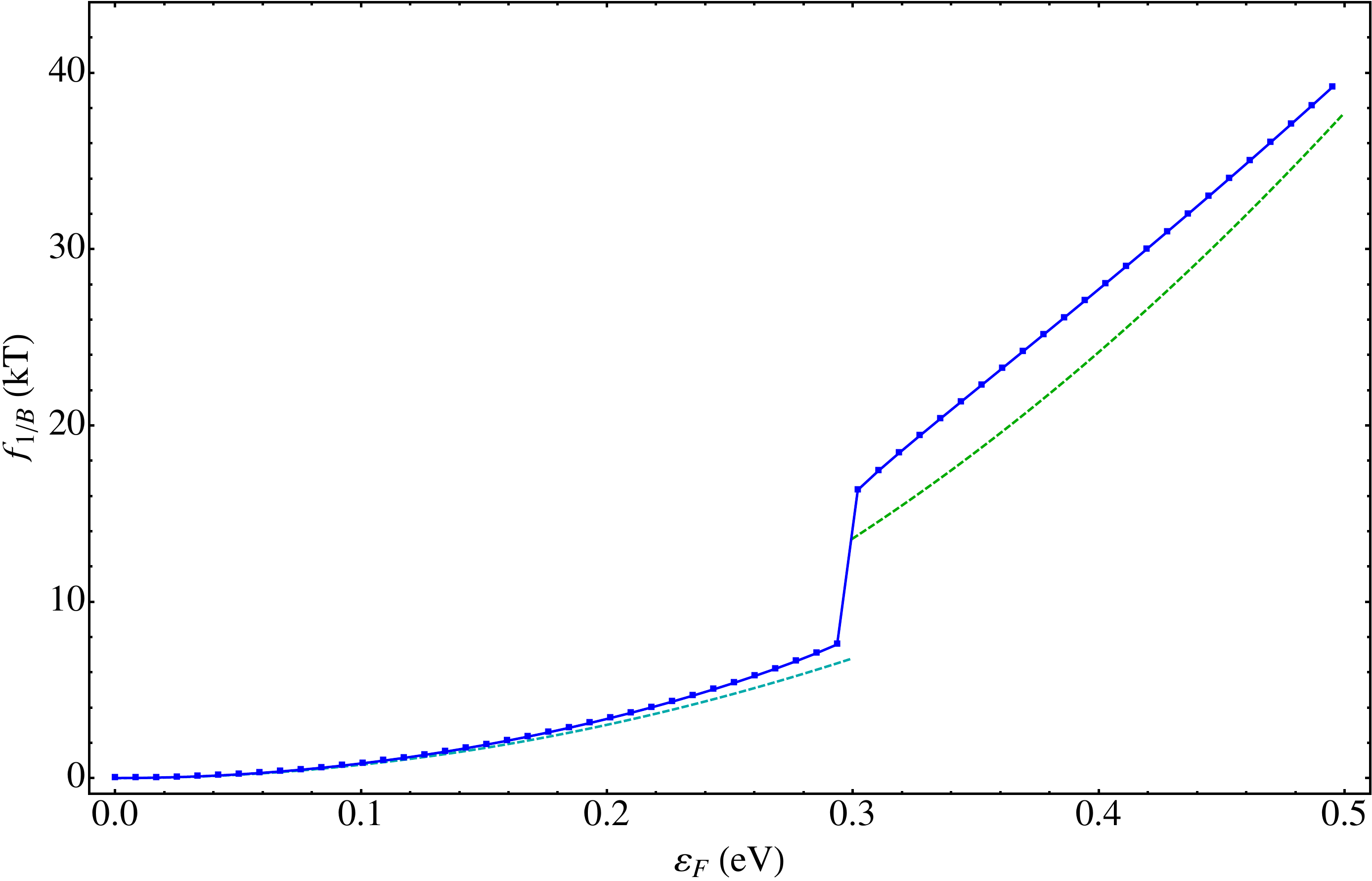}
\caption{Fequency for magnetic fields applied along the $z$-direction as a function of the energy is plotted here. The solid curve shows the result corresponding to the only extremal orbit in this case. Note the orders of magnitude difference compared to fig. \ref{fig:planarFreq} arising from the shape of the torus.}
\label{fig:growthFreq}
\end{figure}

\begin{figure}[ht]
\includegraphics[width=1.02 \linewidth]{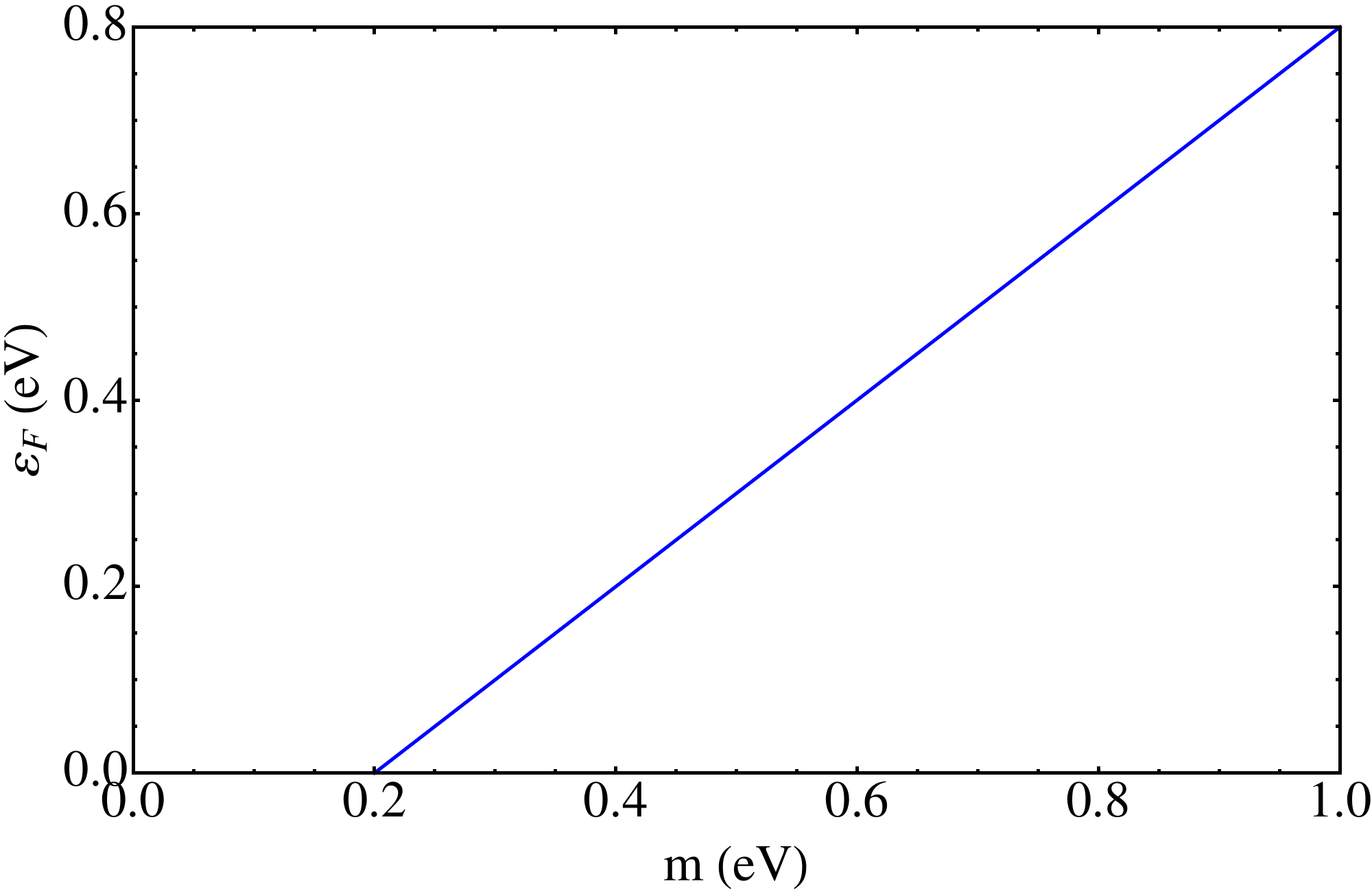}
\caption{As $m$ is varied, the density at which the jump in frequency occurs is modified. Here we plot the variation and note that it occurs for small densities when $m$ is not too much larger than the difference $|\Delta_{S}-\Delta_{D}|$. }
\label{fig:jumpFreq}
\end{figure}

This frequency-doubling is an interesting diagnostic of the line node. The experimental observation of the phenomenon depends on three conditions being satisfied: 1) the ability to tune the density of  electronic carriers in the device; 2) the shape anisotropy of the magnetic insulator being sufficiently strong so as to allow for oscillations to be observed without reorientation of magnetization in the external magnetic field; and  3) the doubling occurs in the low energy regime of the device. While the first two are material challenges, the last can be addressed by looking at the energy at which the doubling occurs as a function of magnetization. From fig. \ref{fig:jumpFreq} it is clear that a parameter regime exists where the critical energy is small, i.e. less then an eV.

\section{Experimental outlook}

Over the last few years remarkable progress has been made in realizing various elements required for the multilayer device. Given the wealth of novel phenomena expected with symmetry-broken surface states \cite{qi, jwu1, zhangrevmodphys, franz1}, detailed theoretical studies have identified candidate materials to activate the time reversal breaking\cite{luoqi}.  On the experimental side a number of ferromagnetic insulators have been grown with the aim of opening a gap in the spectrum of the surface states of topological insulators. Exchange coupling induced symmetry breaking has been observed when Bi$_{2}$Se$_{3}$ is grown on ferromagnetic EuS\cite{moodera1}. The induced magnetic moment at the interface at low temperatures is 1.3 $\pm$ 0.5 $\times$ $10^{2}$ $\mu_{B}/$nm$^{2}$ with a transition at about 20 K.  EuO is a viable candidate but so far only growth on graphene has been demonstrated with a transition temperature of 69 K \cite{swartz}. Cr$_{2}$Ge$_{2}$Te$_{6}$ has a transition temperature of 61 K \cite{cboa} and is another possible substrate for epitaxial growth\cite{petta}. While YIG is an actively researched ferromagnetic insulator, its transition temperature of 559 K results in a constant magnetization at low temperatures which prevents its use as a tunable knob.

An alternative scenario is to follow the original suggestion of [\onlinecite{bb,bhb}] where a magnetically doped topological insulator is sandwiched between normal insulator layers. The ability to magnetically dope topological insulators has been experimentally demonstrated and resulted in the observation of the quantum anomalous hall effect \cite{czchang}.  The temperatures at which the phenomenon is observed is 30 mK while the  Curie temperature is 15 K.  Ordering the moments in plane, rather than perpendicular to the interface will achieve the required geometry. Whether the requirements for line nodal semimetals are as stringent in terms of temperature is yet to be determined. Nevertheless the progress suggests that the prospect of growing devices with Weyl semimetallic characteristics is indeed promising, opening the possibility of new tunable devices discussed in this paper.

\section{Conclusions} 

In this paper we have focussed on the tunability of the low energy sector of a heterostructure which is in a topological semimetallic phase with line nodes. The key insight is the dependence of the dispersion on the strength of the time reversal breaking. For an insulating magnetic layer, this is controlled by the magnetization which in turn depends on temperature. For example EuO has a $T_{c}$ of 69.3 K and cooling provides a knob to continuously vary the magnetization. As one increases the magnetization, the gap in the spectrum closes and the line node appears. This evolves from a closed loop to two open lines which span the Brillouin zone in the direction parallel to the growth axis of the multilayer. The associated changes on density of states and topology of equal energy surfaces results in measurable signatures in thermodynamic and transport properties. We show that the slope of the density of states rises monotonically as a function of $m$ as long as the line node is closed and is roughly constant for larger values. The trend is also reflected in conductivity. 

The toroidal topology also has interesting implications for quantum oscillations in this device. On doping the system the minor radius of the torus grows and for sufficiently large densities the equal energy surface changes topology to a sphere. This is accompanied by the doubling of the oscillation frequency for magnetic fields perpendicular to the symmetry axis of the torus. This is accessible even for small densities. 

As noted in [\onlinecite{bhb}], the nodal line is not robust and that perturbations, such as particle-hole asymmetry, induce an energy dependence to the line where the bands touch. Thus the system is converted to a normal semimetal with electron-hole pockets. Nevertheless the size of these pockets depend on the size of the line node which in turn depends on $m$. Thus the qualitative feature of the evolution of density of states and associated properties as a function of magnetization continue to hold. The same is true of the change in geometry of the nodal line from a closed to an open one. 

VA acknowledges the support of the Hellman foundation.


\end{document}